\documentclass{PoS}

\title{Irradiated shocks in the W28~A2 massive star-forming region: a site for cosmic rays acceleration ?}

\ShortTitle{Cosmic rays acceleration in the shocks of the W28~A2 massive star-forming region ?}

\author{\speaker{Antoine Gusdorf}\\
        LERMA, UMR 8112 du CNRS, Observatoire de Paris, \'Ecole Normale Sup\'erieure, 24 rue Lhomond, F75231 Paris Cedex 05, France\\
        E-mail: \email{antoine.gusdorf@lra.ens.fr}}

\author{Alexandre Marcowith\\
        Laboratoire de Physique Th\'eorique et d'Astroparticules (LPTA), UMR 5207 CNRS, Universit\'e Montpellier II, 13, Place E. Bataillon, 34095 Montpellier Cedex 5, France\\
        E-mail: \email{almarcowith@gmail.com}}

\author{Maryvonne Gerin\\
        LERMA, UMR 8112 du CNRS, Observatoire de Paris, \'Ecole Normale Sup\'erieure, 24 rue Lhomond, F75231 Paris Cedex 05, France\\
        E-mail: \email{maryvonne.gerin@lra.ens.fr}}

\author{Rolf G\"usten\\
         Max Planck Institut f\"ur Radioastronomie, Auf dem H\"ugel 69, 53121 Bonn, Germany\\
        E-mail: \email{rguesten@mpifr.de}}

\abstract{The formation of massive stars play a crucial role in galaxies from numerous points of view. The protostar generates a strong ultraviolet radiation field that ionizes its surroundings, and it drives powerful shock waves in the neighbouring medium in the form of jets and bipolar outflows, whose structure can be partially organized by local, strong magnetic field. Such an ejection activity locally modifies the interstellar chemistry, contributing to the cycle of matter. It also significantly participates to the energetic balance of galaxies. In the latter stages of massive star formation, the protostar is surrounded by an ultra-compact HII region, and irradiates its bipolar outflows, where an intrinsically strong magnetic field structure is associated to the generally high densities. In the HII region, or in the bipolar outflows, the question of \textit{in situ} cosmic rays acceleration can then be raised by the simultaneous presence of strong magnetic fields, significant ionization of the matter, and mechanical energy available in large quantities. In this contribution, we will only summarise the results of our study of potential \textit{in situ} cosmic rays acceleration the irradiated shocks in the W28~A2 massive star forming region, based on observations from the APEX, IRAM 30m, and \textit{Herschel} telescopes, and presented in Gusdorf et al., recently submitted to A\&A.}

\FullConference{Cosmic Rays and the InterStellar Medium - CRISM 2014,\\
		24-27 June 2014\\
		Montpellier, France}

\begin{document}

\section{Introduction}
\label{sec:intro}

One of the least documented aspects of the massive star formation feedback is the role they might play in accelerating cosmic rays (CRs). Indeed, the primary sites for CR acceleration mechanisms have long been considered to be shocks associated to rather young supernova remnants (SNRs, e.g. \cite{Blandford87}). However, \cite{Montmerle79} was the first to seek a correlation between star formation (traced by OB associations) and $\gamma$-ray emission as detected by the Cos B satellite. In this scenario, the OB association served the purpose of low-energy particle injection, and the neighbouring, expanding supernova remnant (SNR) provided the necessary energy to accelerate them. In turn, \cite{Voelk82} investigated the acceleration of cosmic-rays at the terminal shocks of OB star winds, \cite{White85} studied the subsequent synchrotron emission due to particle acceleration in hot stars environments, and \cite{Pollock87} showed evidence for non thermal phenomena in Wolf-Rayet stars at X- and $\gamma$-ray frequencies. Finally, \cite{Romero99} demonstrated a significant evidence for an association of $\gamma$-ray sources detected by EGRET with OB-star forming regions (SFRs). High-energy telescopes were then turned towards massive SFRs, leading to the detection of TeV radiation with HEGRA in the vicinity of the Cygnus OB2 association region (\cite{Aharonian02}), and above 380~GeV with HESS in the young stellar cluster Westerlund 2 (\cite{Aharonian07}). More recently, the detection of TeV emission with \textit{Fermi} was reported in the massive SFR W43 by \cite{Lemoinegoumard11}, and in the Cygnus X superbubble by \cite{Ackermann11}, more or less simultaneously to the study of the association of \textit{Fermi} sources with massive young Galactic objects by \cite{Munaradrover11}. From the modelling point of view, in the massive SFR IRAS 16547--4247, the observed spectral index was found to be consistent with the existence of an uncooled population of relativistic electrons produced by diffusive shock acceleration at a strong non-relativistic shock. This triggered the study of \cite{Araudo07} and \cite{Boschramon10}, who concluded to the possible detection in this region of signatures of accelerated CRs by the Cherenkov arrays such as HESS-II at very high energies ($E_\gamma \gtrsim 100$~GeV), in an affordable amount of time. An overall conclusion is that it is possible that all favourable conditions for CR acceleration be co-existing in massive star-forming regions, where strong magnetic field, a high degree of ionization, and mechanical energy in the form of shocks are simultaneously present, precisely owing to the effects of shocks and HII region. In this contribution, we will only summarise the results of our study of the irradiated shocks in the W28~A2 massive star forming region, based on observations from the APEX, IRAM 30m, and \textit{Herschel} telescopes, and presented in Gusdorf et al., recently submitted to A\&A.

\section{The W28~A2 complex star-forming region}
\label{sec:w28a2}

\begin{figure}
\centering
\includegraphics[width=0.45\textwidth]{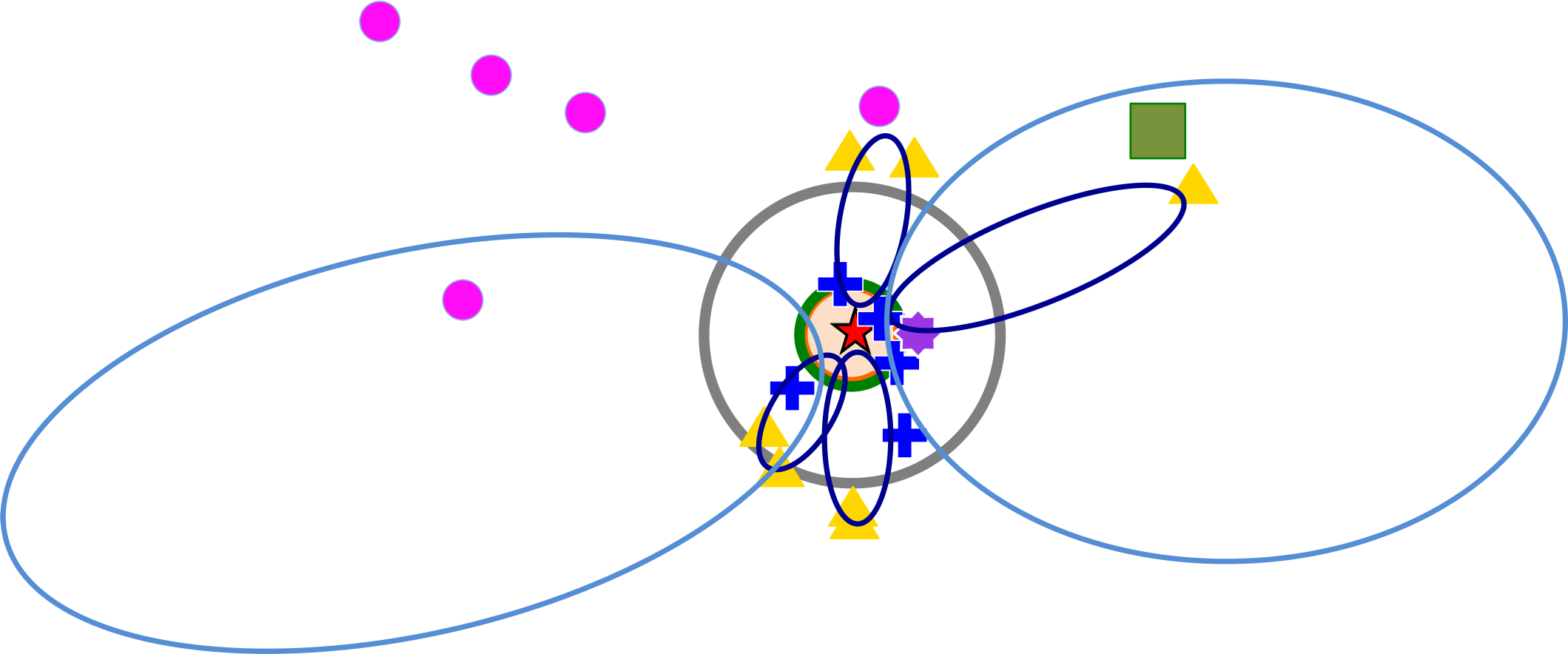}
\caption{A symbolic sketch of the W28 A2 region. The grey circle marks the beam of our analysis. The red star is the central, ionizing, so-called Feldt's star. It is surrounded by an orange circle that delimitates the extent of the ultra-compact HII region, itself surrounded by a dusty and molecular shell shown in the form of a green circle. The blue crosses indicate the positions of five continuum sources. The purple star shows the centre of a Br$\gamma$ outflow. The yellow triangles indicate the H$_2$ knots attributable to outflows apices. The two corresponding outflows are shown in the form of dark blue ellipses. The spectacular outflow observed by \cite{Harvey88} is shown as light blue ellipses. Finally, maser spots are indicated:~NH$_3$ (green squares), Class I CH$_3$OH (pink circles). Not shown are the H$_2$O and OH ones. Adapted from \cite{Hunter08}, shown in Gusdorf et al., subm.} 
\label{figure1}
\end{figure}

W28 A2 is a high-mass star forming site of complex morphology. Its structure and interstellar content is best described by the most recent observations (e.g. \cite{Hunter08}), which benefited from better sensitivity and/or spatial resolution than their older counterparts (e.g. \cite{Harvey88}). A sketch of the region can be found in Figure~\ref{figure1}, indicating the main components of the region:~the central star or star cluster, the surrounding HII region and associated dense cocoon of gas and dust, outflows revealed in various tracers, maser spots, X- and $\gamma$-ray emission peaks. We adopted the distance determination of 1.28$^{+0.09}_{-0.08}$~kpc by~\cite{Motogi11}, obtained through astrometric observations of the 22~GHz water maser with VERA. At the centre of the region, an ionizing star was directly imaged by~\cite{Feldt03}. Given the larger distance these authors associated to this source, \cite{Motogi11} revised its classification in favour of a O8--O8.5 ZAMS. Around this ionizing star, \cite{Hunter08} report the detection of five dust-emission sources, that could be protostars or protostellar cores at various stages of evolution. The central star has carved a roughly circular ultra-compact HII region (e.g., \cite{Moorwood83}, \cite{Wood89}) surrounded by a dense cocoon of dust and molecular gas (e.g., \cite{Su09}). The central region resembling a cluster of sources, drives a gigantic East-West outflow first reported by~\cite{Harvey88}. Other outflows have been identified in the region: two small, nearly North-South ones were detected through interferometric observations of CO and SiO lines (e.g.~\cite{Watson07, Hunter08, Su12}), and their apices could coincide with the observations of H$_2$ knots at 2.12$\mu$m by \cite{Puga06}. Finally, a small outflow has been detected west of the ionizing, central star, in the Br$\gamma$ line by~\cite{Puga06}. The presence of numerous maser spots has also been reported in the region in various species, in OH, H$_2$O, CH$_3$OH, and NH$_3$ (see e.g. \cite{Hunter08} and references therein). \cite{Fish06} and \cite{Stark07} used their OH maser observations to constrain the local strength of the magnetic field to around 2~mG. This value was also measured on a region of large spatial extent ($\gtrsim$10000~A.U.) by the application of the Chandrasekhar-Fermi method to the high angular resolution SMA observations of the polarised dust continuum emission at 870~$\mu$m by \cite{Tang09}. 

W28~A2 has also been widely observed in the high-energy regime owing to its proximity with the W28 SNR, and detected independently from it. Although not seen in the X-ray \textit{ROSAT} and \textit{ASCA} detector ranges (respectively 0.5--2.4 and 0.6--10~keV) by \cite{Rho02}, the HII region was comprised in the peak of emission detected by XMM-Newton, and was very clearly detected by \textit{Chandra}, according to the preliminary results presented by \cite{Rowell10}. Furthermore, its location seems to match that of one very high energy ($E >$ 0.1~TeV) $\gamma$-ray emission peak described in \cite{Aharonian08}:~HESS J1800-240B, suggesting that CR acceleration might be taking place. This view is supported by the most recent $\gamma$-ray observations of the region by AGILE ($E >$ 400~MeV, \cite{Giuliani10}), and by the \textit{Fermi} LAT (from 0.2 to 100~GeV, \cite{Abdo10}), which both confirmed prominent emission on this location. The \textit{Fermi} and HESS combined $\gamma$-ray spectra are accounted for by \cite{Gabici09}, \cite{Li10} and \cite{Hanabata14}, who argued that the strong energetic $\gamma$-ray emission is caused by the collisions from energetic protons (escaping from the SNR shock front) on the W28 A2 molecular cloud, due to the close distance between the high-mass SFR and the W28 SNR. However, the latest study recognized the lack of models to evaluate the possible \textit{in situ} acceleration of CRs within the HII region or the outflows associated to this SFR.

\section{Evidence for irradiated shocks in W28~A2}
\label{sec:irradiated}

\begin{figure}
\centering
\includegraphics[width=\textwidth]{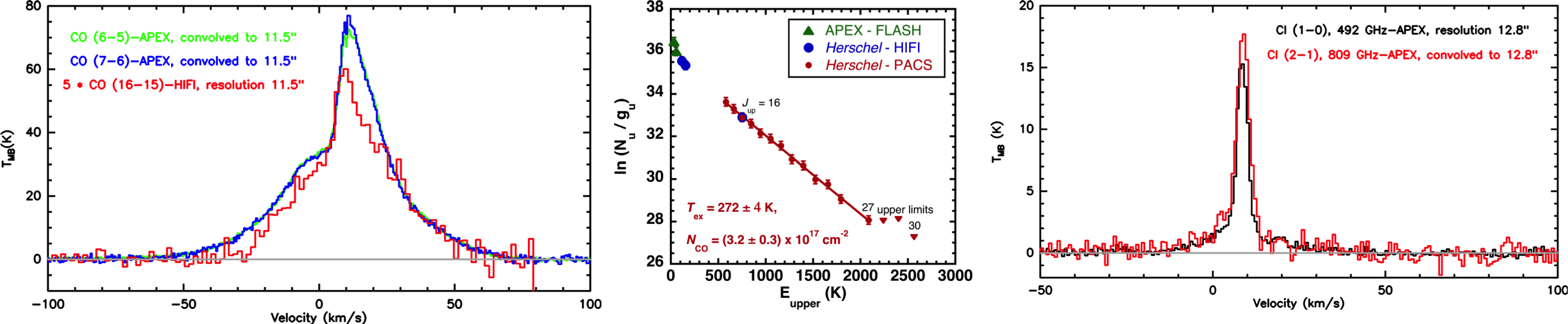}
\caption{\textit{Left panel:} a sample of our CO lines observations: (6--5) and (7--6) (green and blue lines, APEX telescope), (16--15) (red line, \textit{Herschel} telescope). \textit{Central panel:} the CO rotational diagram, obtained from \textit{Herschel} observations (red points). \textit{Right panel:} atomic carbon transitions observed with the APEX telescope: $^3$P$_1$--$^3$P$_0$ and $^3$P$_2$--$^3$P$_1$ (black and red lines). All data were obtained in the central position of W28~A2 (see Figure~1), and extracted from Gusdorf et al., subm.} 
\label{figure2}
\end{figure}

We mapped the emission of the rotational transitions CO with $J_{\rm up}$ = 3, 4, 6, 7 with the APEX telescope, performed pointed observations of the $J_{\rm up}$ = 6, 7, 16 with the HIFI receiver onboard the \textit{Herschel} telescope, and obtained a footprint of the full \lq Spectral Line Energy Distribution' (SLED) accessible to the PACS receiver onboard the \textit{Herschel} telescope. We also observed the $^{13}$CO isotopologue's transitions $J_{\rm up}$ = 3 and 6 with the APEX telescope, while the $^{13}$CO (14--13) line was also detected with PACS. A comparison between the  $J_{\rm up}$ = 6, 7 (from APEX) and 16 (from HIFI) spectra in the central region, at the same spatial resolution, is given in Figure~\ref{figure2}. The presence of CO shocked gas can be seen in these spectra through the line wings extending up to $\pm$80~km~s$^{-1}$ towards the blue-and red-shifted directions. From these spectra, we also inferred that approximately one third of the gas is comprised in each kinematical component: blue-shifted (-100 to 4~km~s$^{-1}$), ambient (4 to 16.5~km~s$^{-1}$), and red-shifted (16.5 to 100~km~s$^{-1}$). From our maps, we also concluded that emission in all CO lines was filling the beam of our observations (12.5$''$, or 0.08~pc); we hence adopted a filling factor of 1 for this emission. However, our observations from the ground also showed that all lines at least up to $J_{\rm up}$ = 6 were optically thick, even in the wings. Our PACS observations, on the other hand, demonstrated that the CO (14--13) line is optically thin, which we extrapolated to the emission from the higher-lying transitions. Using those lines, we could then resort to a rotational diagram approach (shown in Figure~\ref{figure2}), and derived the excitation conditions for CO: the column density in each kinematical component was found to be of the order of 10$^{17}$~cm$^{-2}$, with a minimum kinetic temperature of the order of 270~K.

Similarly, we mapped the $^3$P$_1$--$^3$P$_0$ and $^3$P$_2$--$^3$P$_1$ line emission of [CI] with APEX, with their line wings also broadly filling a beam of 12.5$''$. However, these wings were found to be much less bright than CO ones, as can be seen in Figure~\ref{figure2}. We performed an LVG analysis to infer a column density of (1.5--3)$\times 10^{17}$~cm$^{-2}$ in each blue- and red-shifted components, with respective kinetic temperatures of 140--280 and 55--70~K. These latter values are subject to the calibration uncertainty on the (2--1) line wing emission. 

\begin{figure}
\centering
\includegraphics[width=\textwidth]{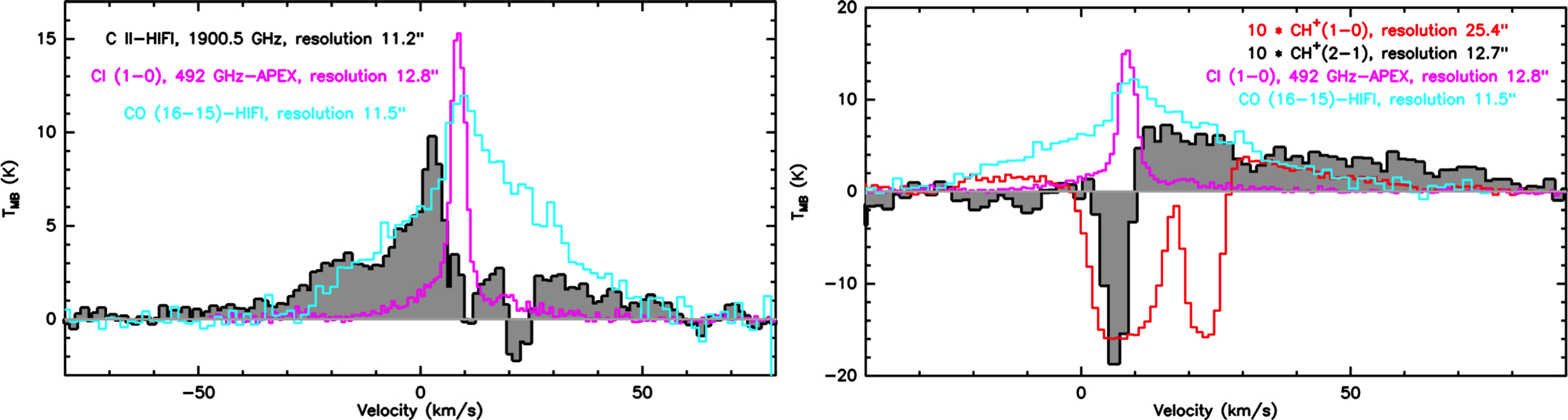}
\caption{\textit{Left panel:} the C$^+$ spectrum (black line and grey histograms). \textit{Right panel:} the CH$^+$ spectra: 1--0 (red line), and 2--1 (black line and grey histograms). All [CII] and CH$^+$ spectra are overlaid with the CO (16--15) and [CI] $^3$P$_1$--$^3$P$_0$ ones from Figure~2 (blue and pink lines); they were obtained with the \textit{Herschel} telescope in the central position of W28~A2 (see Figure~1), and extracted from Gusdorf et al., subm.} 
\label{figure3}
\end{figure}

Additionally, we mapped the [CII] line ($^2$P$_{3/2}$--$^2$P$_{1/2}$) at 1900~GHz with the \textit{Herschel} telescope. Again the emission was found to roughly fill a beam of 12.5$''$. The velocity-resolved profile extracted in the central position shows wings extending up to $\pm$50~km~s$^{-1}$, and constitutes unambiguous evidence that the shocks in the region are irradiated by the UV field of the proto-star(s) (the corresponding spectrum can be seen in Figure~\ref{figure3}). LTE conditions, optical thinness, and an excitation temperature in the range 50--250~K, consistent with the coincidence between [CII] and CO (16--15) line wings were assumed. Our analysis then yielded a column density value of (0.5--1.2)$\times$10$^{18}$~cm$^{-2}$ in the blue-shifted component, and a lower limit of (1.5--3.5)$\times$10$^{17}$~cm$^{-2}$ in the red-shifted component. This result quantitatively supports the view of a significant role of the UV radiation of the forming star, since it makes C$^+$ the dominant form of carbon in the observed shocks. Assuming $n$(C)/$n$(H)=[$n$(CO)+$n$(CI)+$n$(C$^+$)]/$n$(H)=1.4$\times 10^{-4}$, we found an H equivalent column density of (1.1--2.2)$\times$10$^{22}$~cm$^{-2}$ and (0.6--1.1)$\times$10$^{22}$~cm$^{-2}$ in the blue- and red-shifted kinematical components. This allowed us to estimate a lower limit to the ionization fraction in the observed region, given by [$n$(C$^+$)/$n$(H)], and of the order of (0.2--1)$\times10^{-4}$. This range is only a lower limit, as it is based on the assumption that C$^+$ is the dominant charge carrier in the medium. 

This also allowed us to infer the energetic parameters associated to the part of the outflows caught in the beam of our observations, following the method of, e.g. \cite{Bontemps96}. The total outflow mass caught in the beam of our observations was found to be 0.6--1.2~$M_\odot$. The assumed outflow velocity in each blue- and red-shifted directions and associated to most of this mass was 50~km~s$^{-1}$, as observed for C$^+$, corresponding to a dynamical age of 760 years. This resulted in a total (blue- and red-shifted wings) mass-loss rate of (0.8--1.2)$\times 10^{-4}~M_{\odot}$~yr$^{-1}$, a total momentum of (32--62)$M_{\odot}$~km~s$^{-1}$, or a mechanical force of (4--8)$\times10^{-2}M_\odot$ km s$^{-1}$ yr$^{-1}$. Additionnally, we measured that a kinetic energy larger than (1.6--3.1)$\times10^{46}$ erg is dissipated in these outflows, corresponding to more than 175--340~$L_{\odot}$. These values are rather typical of bipolar outflow activity driven by an O-type star or cluster containing one (\cite{Lopezsepulcre09}).

Finally, the irradiated nature of the shocks associated to the bipolar outflows was confirmed by the observation of the (1--0) and (2--1) lines of CH$^{+}$ in single-pointing mode with the HIFI receiver onboard \textit{Herschel} telescope (shown in Figure~\ref{figure3}). Both line profiles exhibit wings extending in the red-shifted direction to 80~km~s$^{-1}$. Because of the different beam sizes of these two transitions, we resorted to a simple LTE and optically thin analysis to yield the column density of (4.4--12.7)$\times 10^{12}$~cm$^{-2}$ in the red-shifted component (corresponding to a fraction of (0.4--2.1)$\times 10^{-9}$ of H equivalent). This result is based on an excitation temperature of 50--250~K compatible with the assumption that CH$^+$ is present in the same gas component as CO, CI and C$^+$, as suggested by the similarity of the wing emission of all their observed transitions. This high value for dense gas-CH$^+$ is in the range of that derived in a similarly massive star forming environment, DR21 by \cite{Falgarone10}, who interpreted it by means of \lq irradiated' shock models. 

The presence of shocks hence large mechanical energy budget on a rather large scale, where the ionization fraction is not negligible and where the magnetic field is intense (\cite{Tang09}) led us to start working on possible \textit{in situ} scenarios of CR acceleration in this region.

\section{\textit{In situ} cosmic rays acceleration in W28~A2}
\label{sec:acceleration}

As we have seen in Section~\ref{sec:w28a2}, the W28~A2 complex is coincident with gamma-ray sources detected by Fermi and HESS. One possible explanation of the gamma-ray signal is that cosmic rays accelerated at the forward shock of the nearby SNR W28 escape and then interact with the dense material associated with W28~A2 \cite{Gabici09, Li10, Hanabata14}. Here, we rather study the possibility that {\it in-situ} acceleration of energetic particles are associated with a {\it part} of the observed gamma-ray radiation. Additionally, we only worked on the evaluation of the maximum particle energies, with views to determine if it could be high enough to at least partly account for the HESS and \textit{Fermi} detections. The work presented in Gusdorf et al., submitted, added to the literature, and specially to those of \cite{Motogi11}, and \cite{Tang09}, allowed us to better understand the complex structure of the W28~A2 region. In particular, we could highlight three regions where particle acceleration could take place:
\begin{itemize}
\item (i) in the irradiated shocks associated to the large-scale bipolar outflows, characterized by a shock velocity $U_{\rm sh}$=50~km~s$^{-1}$, age $t_{\rm age}$=760~years and a radius of 16000~AU with an ionization fraction of $\sim10^{-4}$ (all inferred from our carbon species observations), an ambient density $n_{\rm H} = 10^5$~cm$^{-3}$ and temperature $T$=$100$~K (typical of shocked gas densities and inferred from our observations), a magnetic field $B$=2~mG (\cite{Tang09});
\item (ii) in the HII region, assuming the outflows are launched in this region close to the central star, with the following parameters: shock velocity $U_{\rm sh}$=50~km~s$^{-1}$ and age $t_{\rm age}$=600~years (both inferred from our carbon species observations), an ambient density $n_{\rm H} = 5 \times 10^5$~cm$^{-3}$, a magnetic field $B$=2~mG (\cite{Tang09}), a temperature of $T$=10000~K typical of HII regions, and a radius of 3300~AU (\cite{Motogi11});
\item (iii) in the central cluster, where the supersonic winds of the massive stars can interact and accelerate particles, then injected in the HII region. This scenario is more speculative but might be the most efficient. The central star is of O spectral type, and powers winds with typical terminal velocities $\sim2000-3000$~km~s$^{-1}$. In the following, we use the inner radius of the HII region estimated by \cite{Motogi11} as the typical size: 1200~AU. We also assume an age of 10$^5$~yrs typical of forming stars surrounded by an ultra-compact HII region. Based on a mass-loss rate of $10^{-6} M_{\odot}$~yr$^{-1}$ (more than two orders of magnitude less than the value associated to the bipolar outflow activity and inferred from our observations), the resulting kinetic energy thus deposited by the stellar wind is of the order of 10$^{49}$~ergs.
\end{itemize}

For the scenario (i), we found that CR acceleration was not very likely in the outflow region. It is indeed weakly ionized with a typical ionization fraction $\sim 10^{-4}$. Neutrals can strongly limit the maximum particle energies through ion-neutral damping of the resonantly interacting waves (\cite{Drury96}). We found that ion-neutral collisions are so strong that they can even prevent particle acceleration to enter the relativistic regime. Much faster shocks are needed to generate any $\gamma$-ray signal from these regions. Regarding scenario (iii), inspired by \cite{Parizot04,Cassemontmerle81}, we found that the maximum particle energy is limited by escape losses. We also measured that if a small fraction of the kinetic energy driven by the stellar winds was imparted into energetic particles, it could be sufficient to account for the energy necessary to explain gamma-ray emissions. It is also possible that such a star is surrounded by several less massive ones. In that case, energetic particles can be re-accelerated by multiple shock interactions. We can speculate that an acceleration by the central star/cluster is a possible mechanism of injection of high energy particles in the ultra compact HII region, but this has to be constraint further by additional observations and modelling efforts.

Finally, in the HII region (scenario (ii)), the sonic and Alfvenic shock Mach numbers can be estimated:~$M_{\rm s} \simeq 5.4$ and $M_{\rm a}\simeq 8.1$ respectively, and show that the shocks are strong and super-Alfvenic. The region being fully ionized the maximum particle energy is limited either by the age of the shock or by losses. Particles can be lost for the acceleration process either because they escape upstream once their diffusive length is of the order of a fraction of the shock radius or as they loose their energy radiatively \cite{Berezhko99}. The radiative losses depend on the type of particle: electrons loose their energy either by synchrotron or Inverse Compton losses whereas protons or heavier nuclei loose their energy through inelastic collisions with the surrounding matter. Following the work of \cite{Reynolds98}, we derived the maximum particle energies depending on the dominant energy limiting process. Regarding hadrons, we found that the main limiting factor of the hadron energy in parallel shocks are geometrical losses upstream the shock with typical maximum energies $E_{\rm max, h} \sim Z \times (5-135)$ GeV. As for the case of electrons, the radiative losses strongly limit the maximum electron energies to values $E_{\rm max, e} \sim 5$ GeV. It is not likely to obtain a TeV signal with these particle energies as Inverse-Compton scattering of more energetic photons rapidly fall in the Klein-Nishina regime. We hence overall concluded that unless invoking local magnetic field amplification, faster shocks or/and quasi-perpendicular shock configuration, the maximum hadron energies do not reach the multi TeV energy domain and can hardly explain the HESS signal. However the maximum particle energies are compatible with the Fermi waveband. The low maximum electron energies and the fact that the highest soft photon energies are in the UV band does not favour a leptonic scenario. 


\end{document}